\documentclass[aps,prd,twocolumn,preprintnumbers,superscriptaddress,nofootinbib]{revtex4-1}
\usepackage{graphicx}
\usepackage{epstopdf}
\usepackage{amsmath}
\usepackage{amsfonts}
\usepackage{amssymb}
\usepackage{appendix}
\usepackage{enumerate}
\usepackage{natbib}
\usepackage{comment}
\usepackage{bbold}
\usepackage[shortlabels]{enumitem}
\usepackage{color}
\usepackage{slashed}
\usepackage{subfigure}
\usepackage{setspace}
\usepackage{footnote}
\usepackage{multirow}
\usepackage{booktabs}
\usepackage[colorlinks = true,
            linkcolor = blue,
            urlcolor  = blue,
            citecolor = blue,
            anchorcolor = blue]{hyperref}
\usepackage[capitalize]{cleveref}
\usepackage{braket}
\usepackage{physics}
\usepackage[normalem]{ulem}
\usepackage{url}

\newcommand{\be}{\begin{eqnarray}}
\newcommand{\ee}{\end{eqnarray}}
\newcommand{\ba} {\begin{equation}\begin{aligned}}
\newcommand{\ea} {\end{aligned}\end{equation}}
\newcommand{\bg} {\begin{equation}\begin{gathered}}
\newcommand{\eg} {\end{gathered}\end{equation}}

\newcommand{\beq}{\begin{equation}}
\newcommand{\eeq}{\end{equation}}

\newcommand{\sn}[2]{$#1{\times}10^{#2}$}

\usepackage{orcidlink}

\begin{document}

\title{New Physics in Ultra-High-Energy Neutrino Flavor}

\author{Samiur~R.~Mir\orcidlink{0000-0002-6531-2174}}
\email{samiur.mir@okstate.edu}
\affiliation{Department of Physics, Oklahoma State University, Stillwater, OK 74078, USA}
\author{Carlos A. Argüelles \orcidlink{0000-0003-4186-4182}}
\email{carguelles@fas.harvard.edu}
\affiliation{Department of Physics \& Laboratory for Particle Physics and Cosmology,
Harvard University, Cambridge, MA 02138, USA}
\author{Vedran~Brdar\orcidlink{0000-0001-7027-5104}}
\email{vedran.brdar@okstate.edu}
\affiliation{Department of Physics, Oklahoma State University, Stillwater, OK 74078, USA}

\begin{abstract}
Neutrino flavor measurements test physics beyond the Standard Model that modifies neutrino production, propagation, or detection.
IceCube has measured the flavor composition of the astrophysical neutrino flux up to PeV energies. 
Next-generation experiments will extend flavor measurements to the ultra-high-energy (UHE) regime, $100\,\mathrm{PeV}$--$10\,\mathrm{EeV}$, by combining in-ice radio detection with Earth-skimming tau neutrinos. 
We present the first study of the new-physics reach of these projected UHE flavor measurements. 
We take Lorentz invariance violation (LIV) as a benchmark, since its effects grow as a power of the neutrino energy. 
We find that UHE flavor ratios will be the most sensitive probe of LIV for operators of dimension four and higher, improving on current bounds from all sectors by up to twenty orders of magnitude. 
Three-flavor observables improve this sensitivity by up to an order of magnitude over Earth-skimming tau neutrinos alone.
\end{abstract}

\maketitle
\noindent
\textbf{Introduction}.
The flavor composition of high-energy astrophysical neutrinos is a sensitive probe of new physics~\cite{Bustamante:2010nq,Mehta:2011qb,Arguelles:2015dca,Bustamante:2015waa,Shoemaker:2015qul,Gonzalez-Garcia:2016gpq,Brdar:2016thq,Li:2016kra,Rasmussen:2017ert,Brdar:2018tce,Farzan:2018pnk,Arguelles:2019tum,Ahlers:2020miq,Song:2020nfh,Arguelles:2022tki,Carloni:2022cqz,Dev:2023znd,Agarwalla:2023sng,Arguelles:2024cjj,Dev:2024yrg,Fong:2024msb,Carloni:2025dhv,Bustamante:2026aur}.
The vacuum Hamiltonian scales as $\Delta m^2/E$ and is therefore suppressed at high energy, whereas operators from physics beyond the Standard Model (BSM) typically grow with energy.
Over the cosmological baselines traveled by astrophysical neutrinos, even minute BSM effects accumulate and shift the flavor composition at Earth away from its expected value near $(1:1:1)$.
The flavor composition is conveniently displayed in a ternary plot, the flavor triangle, whose three sides correspond to the electron-, muon-, and tau-neutrino fractions.

IceCube has measured the flavor composition of the astrophysical neutrino flux between TeV and PeV energies~\cite{IceCube:2015rro,IceCube:2015gsk,IceCube:2025ole}.
Charged-current interactions of muon neutrinos produce muon tracks, while electron and tau neutrinos produce cascades; tau neutrinos can further be identified through the double-cascade morphology~\cite{IceCube:2013cdw,IceCube:2014stg,IceCube:2015qii,IceCube:2020wum,IceCube:2020fpi,IceCube:2021rpz,IceCube:2024nhk}.
These measurements already yield the strongest constraints on several classes of Lorentz invariance violation (LIV) operators~\cite{Katori:2016eni,IceCube:2021tdn,Arguelles:2023jkh}.
Their reach is set by the highest neutrino energies observed, since the effect of a dimension-$d$ LIV operator on flavor grows as $E^{d-3}$.

Next-generation experiments will extend flavor measurements to the ultra-high-energy (UHE) regime, from $100\,\mathrm{PeV}$ to $10\,\mathrm{EeV}$, where the flux is expected to be predominantly cosmogenic~\cite{Greisen:1966jv,Zatsepin:1966jv,Halzen:2002pg}.
An array of operating and proposed detectors targets this regime, including ARA~\cite{Allison:2011wk}, ARIANNA~\cite{ARIANNA:2019scz}, Auger~\cite{PierreAuger:2015eyc}, BEACON~\cite{Southall:2022yil}, GRAND~\cite{GRAND:2018iaj}, IceCube-Gen2~\cite{IceCube-Gen2:2021rkf}, POEMMA~\cite{POEMMA:2020ykm}, PUEO~\cite{PUEO:2020bnn}, RNO-G~\cite{RNO-G:2020rmc}, TAMBO~\cite{TAMBO:2025jio}, and Trinity~\cite{Otte:2025dld}.
The most developed detection strategy is the Earth-skimming $\nu_\tau$ channel~\cite{Feng:2001ue,Aramo:2004pr,Reno:2019qmk}, in which a $\nu_\tau$ interacts in the Earth and the emerging tau lepton decays in the atmosphere, initiating an upward-going air shower.
These showers can be detected through their radio emission by GRAND, BEACON, and PUEO; through their optical Cherenkov emission by POEMMA and Trinity; and through the shower particles reaching the ground by Auger and TAMBO.
In-ice radio detectors such as ARA, ARIANNA, RNO-G, and IceCube-Gen2 provide two additional complementary channels~\cite{Coleman:2024scd}: $\nu_e$ charged-current events are identified through the time-extended radio signal of showers elongated by the Landau--Pomeranchuk--Migdal effect~\cite{Landau:1953um,Landau:1965ksp,Migdal:1956tc}; $\nu_\mu$ and $\nu_\tau$ charged-current events can be detected through the multiple secondary showers produced by the outgoing leptons~\cite{Garcia-Fernandez:2020dhb}.
Combining all three channels, Ref.~\cite{Burgos-Mondejar:2026ikk} recently derived the first projections for the UHE neutrino flavor triangle.

In this work, we confront BSM physics with these UHE flavor projections for the first time.
We take LIV as a benchmark, since its power-law energy dependence makes the UHE regime the natural place to search for it.
We find that UHE flavor measurements will be the most sensitive test of Lorentz invariance for operators of dimension four and higher, improving on current bounds from all sectors by up to twenty orders of magnitude.
We further show that having all three flavors available improves the sensitivity by up to an order of magnitude over the Earth-skimming $\nu_\tau$ channel alone.
The same approach applies to any BSM scenario that modifies neutrino production, propagation, or detection, and our results motivate a broader survey of such scenarios with UHE flavor data.

\begin{figure}[t!]
  \centering     
  \includegraphics[width=0.9\linewidth]{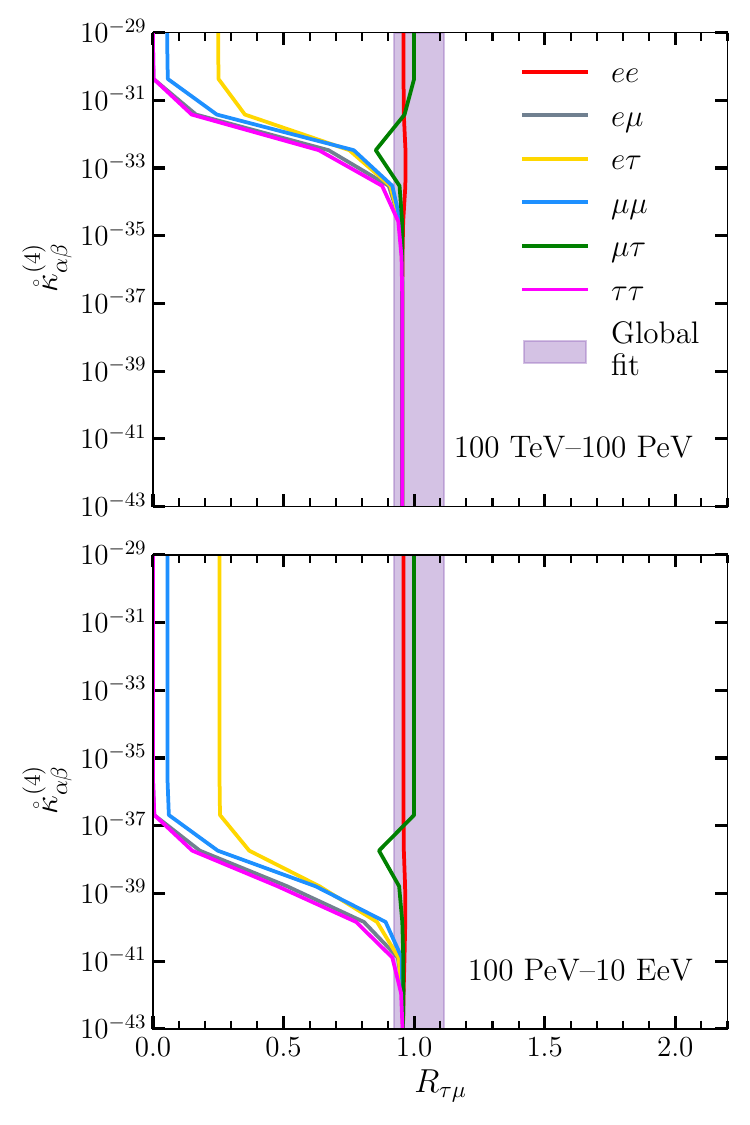}
  \caption{The energy-averaged flavor ratio $R_{\tau\mu}\equiv \phi_\tau/\phi_\mu$ for each of the six flavor combinations of $\mathring{\kappa}^{(4)}_{\alpha\beta}$. In the top (bottom) panel, we consider $E_\text{min}=100\,\mathrm{TeV}$ and $E_\text{max}=100\,\mathrm{PeV}$ ($E_\text{min}=100\,\mathrm{PeV}$ and $E_\text{max}=10\,\mathrm{EeV}$). We also show the preferred region from the global fit~\cite{Wen:2026ngx} as a purple vertical band.
} 
\label{fig:flux_ratio_2D}
\end{figure}

\noindent
\textbf{New Physics in UHE Neutrino Flavor}.
New physics that modifies neutrino propagation can be described by an effective Hamiltonian in the flavor basis,
\begin{align}
    H_{\mathrm{tot}} = H_0 + \sum_n \left(\frac{E}{\Lambda_n}\right)^{n} \tilde{U}_n\, O_n\, \tilde{U}_n^\dagger\,,
    \label{eq:H}
\end{align}
where $H_0$ is the vacuum Hamiltonian, $O_n$ and $\Lambda_n$ set the scale of the new-physics operator of order $n$ in energy, and $\tilde{U}_n$ describes its flavor structure~\cite{Arguelles:2015dca}. Since $H_0\propto 1/E$, any operator with $n\geq 0$ dominates at sufficiently high energy, which is what makes UHE flavor a powerful probe. In this work we take isotropic LIV as a benchmark. The vacuum Hamiltonian is
\begin{align}
    H_0 = U\,\mathrm{diag}\left(0,\frac{\Delta m_{21}^2}{2E},\frac{\Delta m_{31}^2}{2E}\right)U^\dagger,
    \label{eq:H0}
\end{align}
where $U$ is the Pontecorvo--Maki--Nakagawa--Sakata (PMNS) leptonic mixing matrix, $\Delta m^2_{ij}\equiv m_i^2-m_j^2$ is the neutrino mass-squared difference corresponding to $m_i$ and $m_j$, and $E$ is the neutrino energy.

The LIV terms, corresponding to $n=d-3$ in \cref{eq:H}, can be written in terms of flavor-dependent LIV coefficients~\cite{Kostelecky:2011gq,IceCube:2017qyp,Telalovic:2025xor}
\begin{align}
    H_{\mathrm{LIV}} = \sum_{d=2}^{\infty} (-1)^{d-1} E^{d-3} \begin{pmatrix}
\mathring{\kappa}^{(d)}_{ee} & \mathring{\kappa}^{(d)}_{e\mu} & \mathring{\kappa}^{(d)}_{e\tau} \\
\mathring{\kappa}^{(d)}_{\mu e} & \mathring{\kappa}^{(d)}_{\mu\mu} & \mathring{\kappa}^{(d)}_{\mu\tau} \\
\mathring{\kappa}^{(d)}_{\tau e} & \mathring{\kappa}^{(d)}_{\tau\mu} & \mathring{\kappa}^{(d)}_{\tau\tau}
\end{pmatrix}\,, 
\label{eq:HLIV_kappa}
\end{align}
where $\mathring{\kappa}^{(d)}_{\alpha \beta}$ is the isotropic LIV coefficient associated with the operator of dimension $d$.\footnote{Hermiticity of the Hamiltonian implies $\mathring{\kappa}^{(d)}_{\alpha \beta} = \mathring{\kappa}^{(d)\,*}_{\beta \alpha}$, so there are at most six independent LIV coefficients for each dimension $d$. In this work, we only consider real values of $\mathring{\kappa}^{(d)}_{\alpha \beta}$.}

The flavor-transition probability for neutrinos produced with flavor $\alpha$ at redshift $z$ and detected with flavor $\beta$ at $z=0$ in the presence of LIV is given by
\begin{align}
    P_{\alpha\beta}(E,z)=\sum_{i=1}^{3} \left| V_{\alpha i}(E(1+z)) \right|^2 \left| V_{\beta i}(E) \right|^2\,,
    \label{eq:probb}
\end{align}
where $V$ is the mixing matrix obtained by diagonalizing \cref{eq:H}.

Because of the power-law energy dependence in \cref{eq:HLIV_kappa}, nonzero LIV can induce a strong energy dependence in the flavor composition of the neutrino flux at Earth. We therefore define the energy-averaged fraction of neutrinos of flavor $\alpha$ over the energy range of interest as~\cite{Testagrossa:2023ukh,Telalovic:2025xor,Brdar:2026jbu} 
\begin{align}
	f_{\alpha,\oplus} = \frac{\int_{E_\text{min}}^{E_\text{max}} \phi_\alpha(E) \, dE}{\sum_{\beta=e,\mu,\tau} \int_{E_\text{min}}^{E_\text{max}} \phi_\beta(E) \, dE}\,.
	\label{eq:flav_frac}
\end{align}
Here, $\phi_\alpha$ represents the neutrino flux of flavor $\alpha$ at Earth, and $E_\text{min}$ and $E_\text{max}$ determine the considered UHE neutrino energy range. In \cref{eq:flav_frac}, and in general throughout this work, we combine neutrinos and antineutrinos of a given flavor and do not distinguish between them, since their charged-current cross sections become identical at UHE energies~\cite{Gandhi:1995tf,Gandhi:1998ri,Cooper-Sarkar:2011jtt,Connolly:2011vc}. 

To compare with the projections of Ref.~\cite{Burgos-Mondejar:2026ikk}, we adopt $E_\text{min}=100\,\mathrm{PeV}$ and $E_\text{max}=10\,\mathrm{EeV}$ and the three benchmark fluxes used therein, dubbed ``optimistic,'' ``moderate,'' and ``pessimistic.'' In this energy range the flux is dominated by cosmogenic neutrinos, produced in interactions of ultra-high-energy cosmic rays (UHECRs) with the cosmic microwave background (CMB) and extragalactic background light (EBL). Because the flavor-transition probability of \cref{eq:probb} depends on the production redshift, the flux at Earth cannot be taken directly from Ref.~\cite{Burgos-Mondejar:2026ikk}. 
Instead, it must be computed as an integral over the redshift distribution,
\begin{align}
    \phi_\beta(E) = \sum_{\alpha} \int_0^{z_{\max}} dz\, P_{\alpha\beta}(E,z)\, \frac{d\phi_\alpha}{dz}(E,z)\,,
    \label{eq:flux_z}
\end{align}
where $d\phi_\alpha/dz$ is the contribution to the neutrino flux of flavor $\alpha$ at redshift $z$~\cite{Carloni:2025dhv,Brdar:2026jbu}. We obtain $d\phi_\alpha/dz$ by simulating UHECR propagation with \texttt{SimProp-v2r4}~\cite{Aloisio:2017iyh}, with injection parameters fitted to reproduce the three benchmark fluxes; see the End Matter for details. Since the flavor fractions in \cref{eq:flav_frac} are normalized, our results are independent of the overall flux normalization and its uncertainty.

\begin{figure}[t!]
    \centering
       \includegraphics[width=0.95\linewidth]{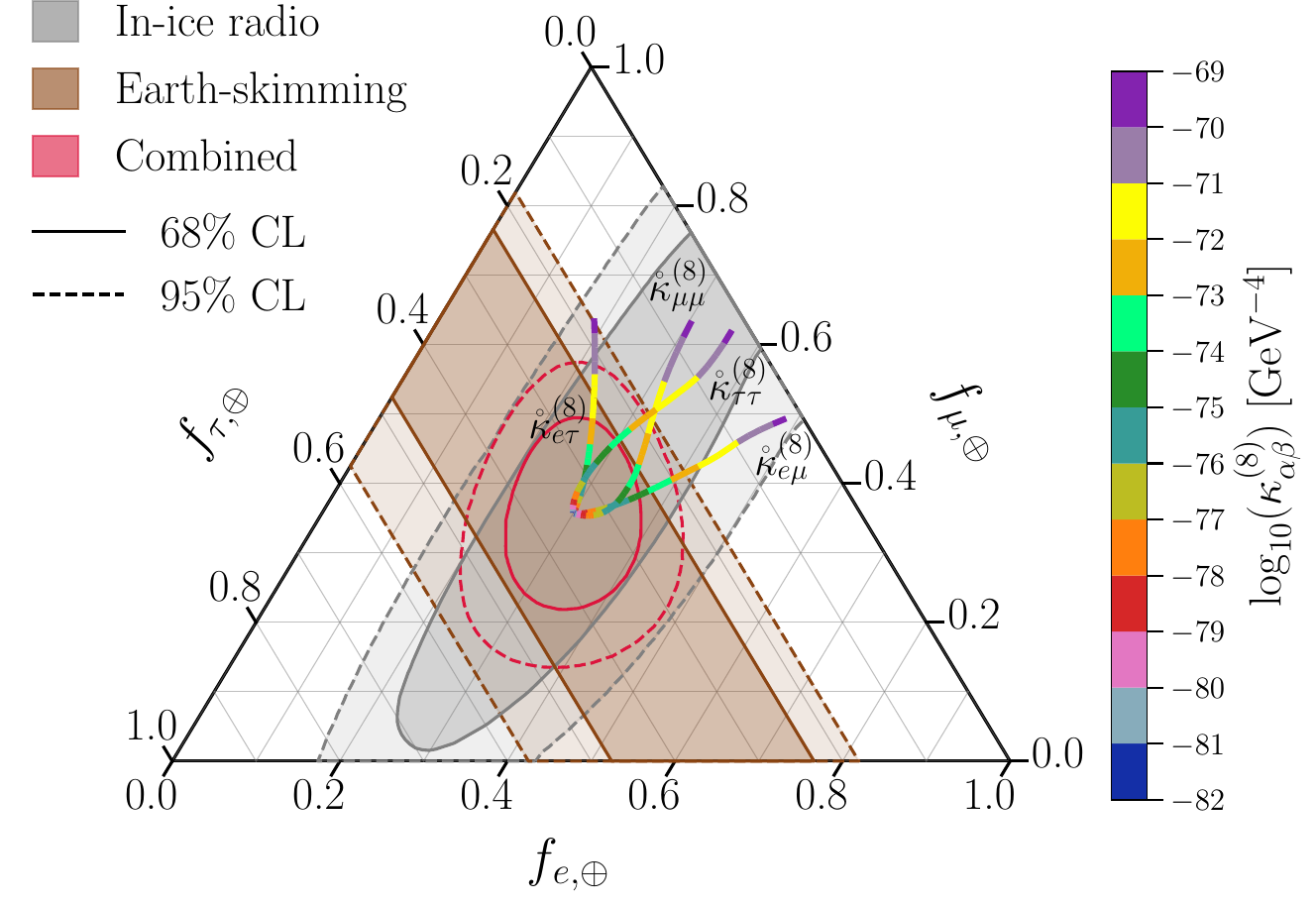}
    \caption{UHE neutrino flavor composition at Earth in the presence of LIV, shown for four different flavor combinations of $\mathring{\kappa}^{(8)}_{\alpha\beta}$, assuming the ``moderate'' flux benchmark. All curves extend toward the upper-right portion of the flavor triangle, with the color bar indicating the corresponding $\mathring{\kappa}^{(8)}_{\alpha\beta}$ values. Also shown are the 68\% and 95\% confidence level (CL) projections for all considered experiments in Ref.~\cite{Burgos-Mondejar:2026ikk} (red), as well as the projections for in-ice radio detection channels (gray) and Earth-skimming tau neutrinos (brown). The projections include GRAND50k alongside the other considered experiments.} 
    \label{fig:ternary}
\end{figure}

\begin{figure*}[t!]
    \centering
       \includegraphics[width=0.95\linewidth]{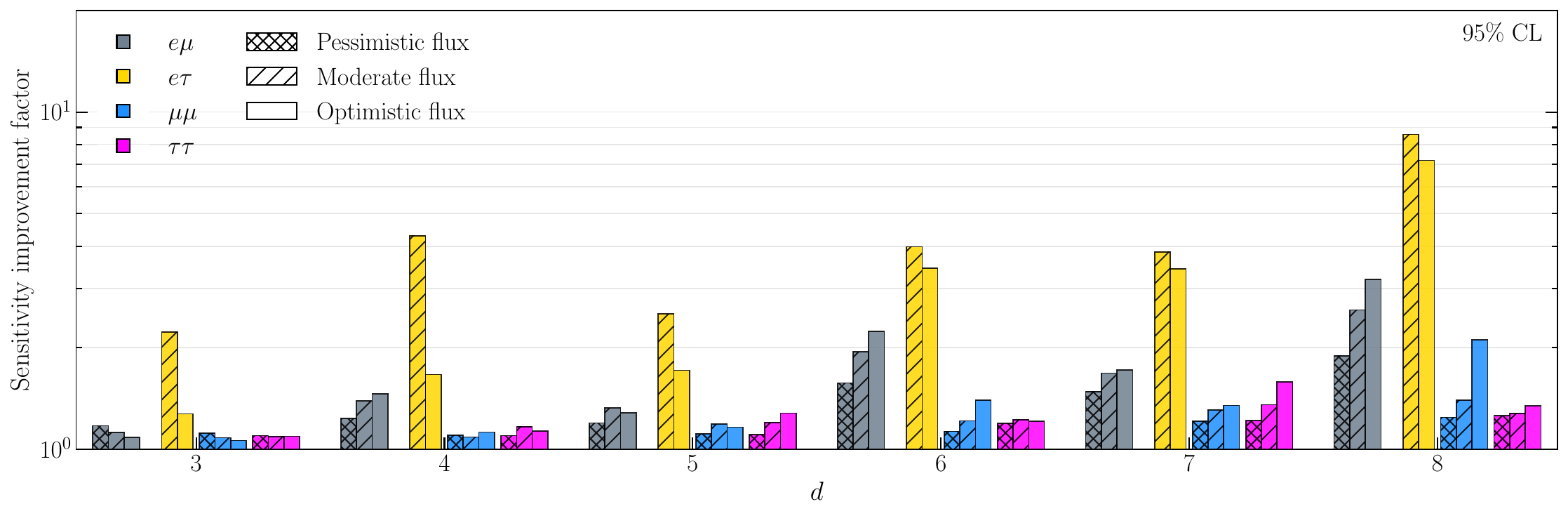}
    \caption{The improvement factors, defined as the ratio of the sensitivity obtained with Earth-skimming tau neutrinos alone to that obtained with the combined analysis including in-ice radio and Earth-skimming tau-neutrino detection, are shown for the GRAND50k configuration, for the three considered UHE neutrino fluxes and for LIV coefficients associated with operator dimensions ranging from $d=3$ to $d=8$ at 95\% CL. For cases in which no ratio is shown, the Earth-skimming analysis does not yield a sensitivity projection.}
    \label{fig:improvements_full}
\end{figure*}
\noindent
\textbf{Results}.
The framework of \cref{eq:H,eq:probb,eq:flux_z} yields the flavor composition at Earth for any new-physics operator that modifies neutrino propagation. We now present results for the LIV benchmark, focusing on minimal scenarios in which a single coefficient $\mathring{\kappa}^{(d)}_{\alpha\beta}$ of a given dimension $d$ is nonzero.

In \cref{fig:flux_ratio_2D}, we present the energy-averaged flavor ratio $R_{\tau\mu}\equiv\phi_\tau/\phi_\mu$ for all six flavor combinations of $\mathring{\kappa}^{(4)}_{\alpha\beta}$. In the top (bottom) panel, we consider the energy range between $100\,\mathrm{TeV}$ and $100\,\mathrm{PeV}$ ($100\,\mathrm{PeV}$ and $10\,\mathrm{EeV}$). In both panels, we generally observe that increasing $\mathring{\kappa}^{(4)}_{\alpha\beta}$ causes the flavor ratio $R_{\tau\mu}$ to deviate from its expected value in the absence of LIV, which is approximately $1$. The exceptions are $\mathring{\kappa}^{(4)}_{\mu\tau}$ and $\mathring{\kappa}^{(4)}_{ee}$, for which we have already demonstrated weak LIV sensitivity in Ref.~\cite{Brdar:2026jbu}. For the other four parameters, deviations from $R_{\tau\mu}\approx 1$ are generally expected to emerge when $H_{\mathrm{LIV}}$ becomes comparable to the vacuum Hamiltonian. In general, this occurs when~\cite{Brdar:2026jbu}
\begin{align}
    \mathring{\kappa}^{(d)}_{\alpha \beta}  \approx  10^{-23} \left(\frac{\mathrm{GeV}}{E}\right)^{d-2}\,\mathrm{GeV}^{4-d}\,.
\label{eq:kappa_compare_LIV_vac}
\end{align}
For $d=4$ and $E=100\,\mathrm{TeV}$ ($E=100\,\mathrm{PeV}$), this yields $\mathring{\kappa}^{(4)}_{\alpha \beta} \approx 10^{-33}$ ($\mathring{\kappa}^{(4)}_{\alpha \beta} \approx 10^{-39}$), which is consistent with the results shown in \cref{fig:flux_ratio_2D} (for completeness, we also present the corresponding results for $R_{e\mu}\equiv\phi_e/\phi_\mu$ in \cref{fig:flux_ratio_2Demu}). This demonstrates that higher neutrino energies can lead to substantially stronger constraints on LIV coefficients, motivating the use of experiments specifically designed to probe the UHE frontier.

While \cref{fig:flux_ratio_2D} illustrates the impact of LIV on a single flavor ratio, $R_{\tau\mu}$, in what follows we consider the full flavor composition of UHE neutrinos. To this end, we focus on ternary plots, with the three sides corresponding to the flavor fractions $f_{e,\oplus}$, $f_{\mu,\oplus}$, and $f_{\tau,\oplus}$ defined in \cref{eq:flav_frac}. In \cref{fig:ternary}, we present the flavor composition of UHE neutrinos in the energy range between $100\,\mathrm{PeV}$ and $10\,\mathrm{EeV}$ for the case where a single $\mathring{\kappa}^{(8)}_{\alpha\beta}$ coefficient is nonzero. We consider all flavor combinations apart from ($e, e$) and ($\mu, \tau$), which have been shown in \cref{fig:flux_ratio_2D} to lead to tiny effects even for large LIV coefficients. The calculation is performed using \texttt{SimProp-v2r4}, with which we interface our calculation of the flavor-transition probabilities in the presence of LIV, as discussed in the previous section. In \cref{fig:ternary}, we consider the ``moderate'' UHE neutrino flux at Earth. We employ NuFIT~\cite{Esteban:2020cvm,Esteban:2024eli} best-fit values for the mixing angles and CP phase, without taking into account the current uncertainties on these parameters. This is because, by the time cosmogenic neutrinos are detected with large statistics by different experiments, we expect the uncertainties on the PMNS matrix elements to be negligible~\cite{Song:2020nfh}, thanks to the lower-energy neutrino experimental program led by DUNE~\cite{DUNE:2015lol}, JUNO~\cite{JUNO:2015zny}, Hyper-Kamiokande~\cite{Hyper-Kamiokande:2018ofw}, the IceCube Upgrade~\cite{Ishihara:2019aao}, and KM3NeT/ORCA~\cite{KM3Net:2016zxf}.  We observe that the curves corresponding to different LIV coefficients in \cref{fig:ternary} extend toward the upper-right portion of the flavor triangle. This behavior is consistent with the broader parameter scan shown in \cref{fig:ternary_scan}, where the green region, corresponding to an initial flavor composition of $f_{e,\mathrm{S}}:f_{\mu,\mathrm{S}}:f_{\tau,\mathrm{S}} \simeq 1/3:2/3:0$ expected for cosmogenic neutrinos, is precisely located in this upper-right portion of the triangle.

The color bar in \cref{fig:ternary} shows the value of the $\mathring{\kappa}^{(8)}_{\alpha\beta}$ coefficient corresponding to the color at each location along the curves in the flavor triangle. These results are compared with the 68\% (solid) and 95\% CL (dashed) UHE neutrino flavor triangle projections from Ref.~\cite{Burgos-Mondejar:2026ikk}, which combine the in-ice radio detectors ARA, ARIANNA, RNO-G, and IceCube-Gen2 with the Earth-skimming experiments TAMBO, POEMMA, Trinity, Auger, and GRAND. Since the sensitivity of GRAND depends strongly on its final size, we consider three configurations, GRAND10k, GRAND50k, and GRAND200k, where the number denotes the number of antennas in the array; all other experiments are kept at their nominal configurations. \Cref{fig:ternary} shows the GRAND50k case, while the results with GRAND10k and GRAND200k are shown in \cref{fig:ternary_10k_200k}. Ref.~\cite{Burgos-Mondejar:2026ikk} presents projections only for GRAND200k; the authors provided us with the sensitivity curves for the other two configurations. The projections using only the in-ice radio channels~\cite{Coleman:2024scd} are shown in gray, and those using only the Earth-skimming experiments in brown.

The intersections of each curve, corresponding to a given flavor combination $(\alpha,\beta)$ of $\mathring{\kappa}^{(8)}_{\alpha\beta}$, with the 68\% (red) and 95\% CL (red dashed) contours determine the sensitivity to the corresponding LIV coefficient. We observe from \cref{fig:ternary} that, for the ``moderate'' UHE neutrino flux, the projected sensitivity to $\mathring{\kappa}^{(8)}_{e\mu}$, $\mathring{\kappa}^{(8)}_{\mu\mu}$, and $\mathring{\kappa}^{(8)}_{\tau\tau}$ is around $10^{-72} \,\text{GeV}^{-4}$  at 95\% CL, while the sensitivity to $\mathring{\kappa}^{(8)}_{e\tau}$ is approximately two orders of magnitude weaker. As far as in-ice radio detection is concerned, the curves associated with all $\mathring{\kappa}^{(8)}_{\alpha\beta}$ generally stay within the gray region, indicating that such detection strategies, when considered on their own, are not very powerful for constraining LIV. This statement does not hold for Earth-skimming tau neutrinos, for which we observe that the $\mathring{\kappa}^{(8)}_{\alpha\beta}$ curves generally overshoot the boundaries of the brown region. Nevertheless, in particular for $\mathring{\kappa}^{(8)}_{e\tau}$, the intersection with the red region occurs for a visibly smaller LIV coefficient, showcasing the advantage of combining all detection technologies for constraining BSM physics.

The improvement factor in sensitivity is obtained by comparing the combined sensitivity including in-ice radio channels and Earth-skimming tau neutrinos with the sensitivity based on Earth-skimming tau neutrinos alone. It is quantified in \cref{fig:improvements_full} for all three considered UHE neutrino fluxes and for LIV coefficients associated with operator dimensions ranging from $d=3$ to $d=8$ at 95\% CL, assuming the GRAND50k configuration.
Overall, we find an improvement in sensitivity of up to an order of magnitude for $\mathring{\kappa}^{(d)}_{e\tau}$, with appreciable effects also observed for $\mathring{\kappa}^{(d)}_{e\mu}$. This clearly demonstrates the advantage of combining all available detection strategies to constrain LIV and, more broadly, probe BSM physics.
For completeness, we also present results for GRAND10k and GRAND200k in \cref{fig:improvements_10k_200k}. For GRAND200k, the improvement achieved by considering all detection channels compared with using tau neutrinos alone is more modest but still reaches 70\% for $\mathring{\kappa}^{(8)}_{e\tau}$. This is because the tau-neutrino detection channel of GRAND200k already provides exceptional sensitivity to UHE neutrinos, thus contributing dominantly to the combined three-flavor sensitivity.

\begin{table}[t!]
\centering
\small
\setlength{\tabcolsep}{4pt}
\renewcommand{\arraystretch}{1.0}
\begin{tabular}{|c|l|c||c|l|c|}
\hline
$d$ & method & limit & $d$ & method & limit \\
\hline\hline
\multirow{2}{*}{3} & CMB~\cite{Caloni:2022xsv} & $10^{-44}$ & \multirow{2}{*}{6} & astro.\ $\nu$~\cite{IceCube:2021tdn} & $10^{-42}$ \\
 & $\bigstar$ & $4\times10^{-31}$ & & $\bigstar$ & $2\times10^{-58}$ \\
\hline
\multirow{2}{*}{4} & GRB~\cite{Kostelecky:2013rta} & $10^{-38}$ & \multirow{2}{*}{7} & astro.\ $\nu$~\cite{IceCube:2021tdn} & $10^{-47}$ \\
 & $\bigstar$ & $2\times10^{-40}$ & & $\bigstar$ & $9\times10^{-67}$ \\
\hline
\multirow{2}{*}{5} & astro.\ $\nu$~\cite{IceCube:2021tdn} & $10^{-37}$ & \multirow{2}{*}{8} & astro.\ $\nu$~\cite{IceCube:2021tdn} & $10^{-52}$ \\
 & $\bigstar$ & $7\times10^{-49}$ & & $\bigstar$ & $3\times10^{-76}$ \\
\hline
\end{tabular}
\caption{Strongest existing limits on LIV coefficients $\mathring{\kappa}^{(d)}_{\alpha\beta}$ for operator dimensions $d=3$--$8$ (order of magnitude), compared with the projected 95\% CL sensitivities of the three-flavor UHE neutrino analysis presented in this work ($\bigstar$). All values in units of $\mathrm{GeV}^{4-d}$.}
\label{tab:lv_limits_short}
\end{table}

For completeness, in addition to the improvement factors, defined as ratios of the corresponding LIV sensitivity values, we present the individual sensitivities entering these ratios in \cref{tab:liv_sensitivity}. Specifically, we report the minimum values of $\mathring{\kappa}^{(d)}_{\alpha\beta}$ that can be probed for all three considered fluxes, both for the combined analysis and for Earth-skimming tau neutrinos alone, across all considered experiments and the three GRAND configurations.

Finally, \cref{tab:lv_limits_short} compares the strongest existing constraints on LIV coefficients for operator dimensions $d=3$--$8$ with the sensitivities obtained in this work, indicated by $\bigstar$ in the table. For each dimension $d$, we report the strongest sensitivity across 
all different flavor combinations of LIV coefficients. Future UHE neutrino measurements could substantially improve the constraints on LIV coefficients for all operator dimensions above $d=3$, reaching improvements of up to $\sim 20$ orders of magnitude for $d=8$. These results demonstrate that UHE neutrino flavor measurements have the potential to revolutionize the search for LIV and physics beyond the Standard Model in general.

\noindent
\textbf{Summary and Conclusions}.
Motivated by recently derived projections for UHE neutrino flavor measurements combining in-ice radio channels and Earth-skimming tau neutrinos, we have investigated the potential of these detection strategies to probe BSM physics by considering the case of Lorentz invariance violation (LIV). We find that UHE flavor measurements will be the most sensitive test of Lorentz invariance for operators of dimension four and higher, improving on current bounds from all sectors by up to twenty orders of magnitude. Having all three flavors available improves the sensitivity to LIV coefficients of dimension $d=3$--$8$ by up to an order of magnitude compared with the Earth-skimming tau-neutrino channel alone. Our approach can be extended to other new-physics scenarios affecting neutrino production, propagation, or detection. Future UHE neutrino observations with sufficiently large statistics and precise flavor measurements could thus open new avenues for probing BSM physics at energies far beyond the reach of terrestrial experiments.

\noindent
\textbf{Acknowledgments}.
We would like to thank Alba Burgos-Mondéjar, Mauricio Bustamante, and Alex Y. Wen for useful discussions. C.A.A. are supported by the Faculty of Arts and Sciences of Harvard University, the National Science Foundation (NSF, CAREER Grant \#2239795), the John Templeton Foundation (Grant \#63651), the Research Corporation for Science Advancement, and the David \& Lucile Packard Foundation. The work of V.B. and S.R.M. is supported by the United States Department of Energy Grant No.~DE-SC0016013.

\section*{End Matter}
\noindent
\textbf{Cosmogenic neutrino flux modeling}.
The three benchmark fluxes of Ref.~\cite{Burgos-Mondejar:2026ikk} share an astrophysical component, a single power law fitted to 9.5~years of IceCube TeV--PeV muon-neutrino data~\cite{Abbasi:2021qfz}, for which we assume the pion-decay source composition $f_{e,\mathrm{S}}:f_{\mu,\mathrm{S}}:f_{\tau,\mathrm{S}}=1/3:2/3:0$~\cite{Beacom:2003nh,Athar:2005wg}. The cosmogenic components of the ``optimistic'' and ``pessimistic'' fluxes are taken from Refs.~\cite{Bergman:2021djm,vanVliet:2019nse}, and the ``moderate'' flux is their geometric mean.

We reproduce these fluxes with \texttt{SimProp-v2r4}~\cite{Aloisio:2017iyh}, which propagates UHECRs injected at redshifts up to $z_{\max}=4$ through the CMB and EBL and records the production redshift and energy of each secondary neutrino. We vary the injected UHECR mass composition, the injection spectral index, and the source emissivity, and fit them to the benchmark fluxes; the resulting parameters are listed in \cref{tab:1}.

The maximum energy, $E_{\mathrm{max}}^{\mathrm{CR}}$, of UHECR species with atomic number $Z$ is determined by the maximum rigidity $R_{\mathrm{max}}$, such that $E_{\mathrm{max}}^{\mathrm{CR}}=Z R_{\mathrm{max}}$. For all three benchmark fluxes, MC events are generated by injecting UHECRs with an $E^{-\gamma_{\rm inj}}$ spectrum up to a maximum redshift of $z=4$. The events are subsequently weighted to reproduce the physical UHECR spectrum, characterized by the spectral index $\gamma_{\rm sp}$. The source redshift evolution is parametrized as $(1+z)^3$ for $z<1.5$ and $2.5^3$ for $z>1.5$~\cite{vanVliet:2019nse}. The EBL model, implemented in \texttt{SimProp-v2r4} to produce neutrinos, is taken from Gilmore et al.~2012~\cite{Gilmore:2011ks}. 

\begin{figure}[h!]
    \centering
   \includegraphics[width=0.95\linewidth]{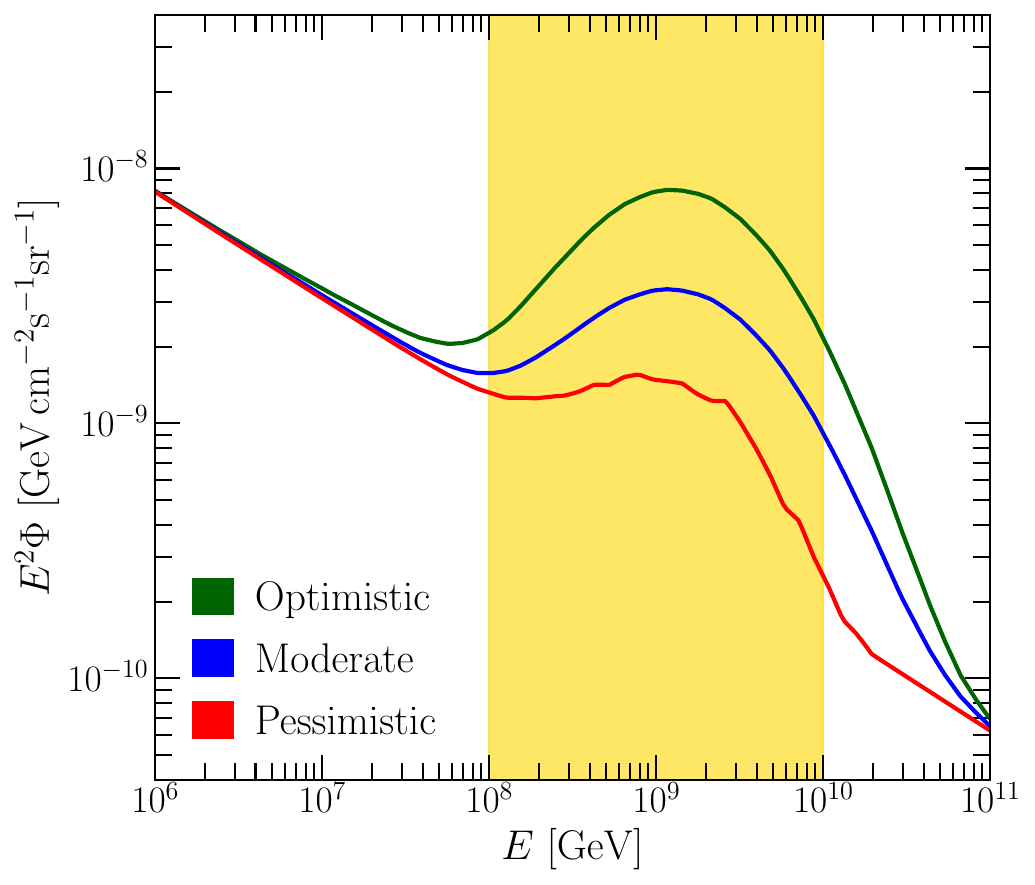}
   \caption{The sum of the astrophysical and cosmogenic components of the three benchmark neutrino fluxes. The astrophysical component is modeled with a single power law, while the cosmogenic components are obtained by fitting UHECR parameters to reproduce the fluxes from Ref.~\cite{Burgos-Mondejar:2026ikk} at Earth. The energy range of interest, between $100\,\mathrm{PeV}$ and $10\,\mathrm{EeV}$, is highlighted in yellow.} 
    \label{fig:flux}
\end{figure}

The fitted fluxes are presented in \cref{fig:flux}. The flavor-transition probability of \cref{eq:probb} is then applied event by event according to \cref{eq:flux_z}, following Ref.~\cite{Brdar:2026jbu}.

\begin{table}[h!]
\footnotesize
\centering
\setlength{\tabcolsep}{2.4pt}
\renewcommand{\arraystretch}{1.15}
\begin{tabular}{|c|c|c|c|c|c|}
\hline
\begin{tabular}[c]{@{}c@{}}$\nu$ Flux \\label\end{tabular} &
$\gamma_{\rm sp}$ & $\gamma_{\rm inj}$ &
\begin{tabular}[c]{@{}c@{}}$R_{\max}$\\ {[EV]}\end{tabular} &
Composition &
\begin{tabular}[c]{@{}c@{}}Source emissivity\\ $\mathcal{L}_0$\\ {[erg\,Mpc$^{-3}$\,yr$^{-1}$]}\end{tabular} \\
\hline\hline
optimistic~\cite{Bergman:2021djm} & 2.06 & 1 & 3160 &
\begin{tabular}[c]{@{}c@{}}99.2\% He\\ + 0.8\% Fe\end{tabular} & $6.67\times10^{44}$ \\
\hline
moderate & 2.06 & 1 & 3160 &
\begin{tabular}[c]{@{}c@{}}95\% He\\ + 5\% Fe\end{tabular} & $2.60\times10^{44}$ \\
\hline
pessimistic~\cite{vanVliet:2019nse} & 2.0 & 2.0 & 100 &
\begin{tabular}[c]{@{}c@{}}10\% of\\ pure-proton\\only\end{tabular} & $1.34\times10^{45}$ \\
\hline
\end{tabular}
\caption{Parameters employed to generate the three benchmark cosmogenic neutrino fluxes. }
\label{tab:1}
\end{table}


\bibliographystyle{JHEP}
\bibliography{refs}

\clearpage
\newpage
\onecolumngrid

\setcounter{equation}{0}
\setcounter{figure}{0}
\setcounter{table}{0}
\setcounter{page}{1}
\setcounter{section}{0}

\makeatletter
\renewcommand{\thesection}{S\arabic{section}}
\renewcommand{\theequation}{S\arabic{equation}}
\renewcommand{\thefigure}{S\arabic{figure}}
\renewcommand{\thetable}{S\arabic{table}}

\renewcommand{\theHfigure}{S\arabic{figure}}
\renewcommand{\theHtable}{S\arabic{table}}
\renewcommand{\theHequation}{S\arabic{equation}}
\makeatother

\begin{center}
    \textbf{\large Supplemental Material:\\ New Physics in Ultra-High-Energy Neutrino Flavor} \\ 
    \vspace{0.5cm}
    {Samiur~R.~Mir, Carlos~A.~Arg\"uelles, Vedran~Brdar}
\end{center}




\begin{figure}[h!]
  \centering     
  \includegraphics[scale=0.7]{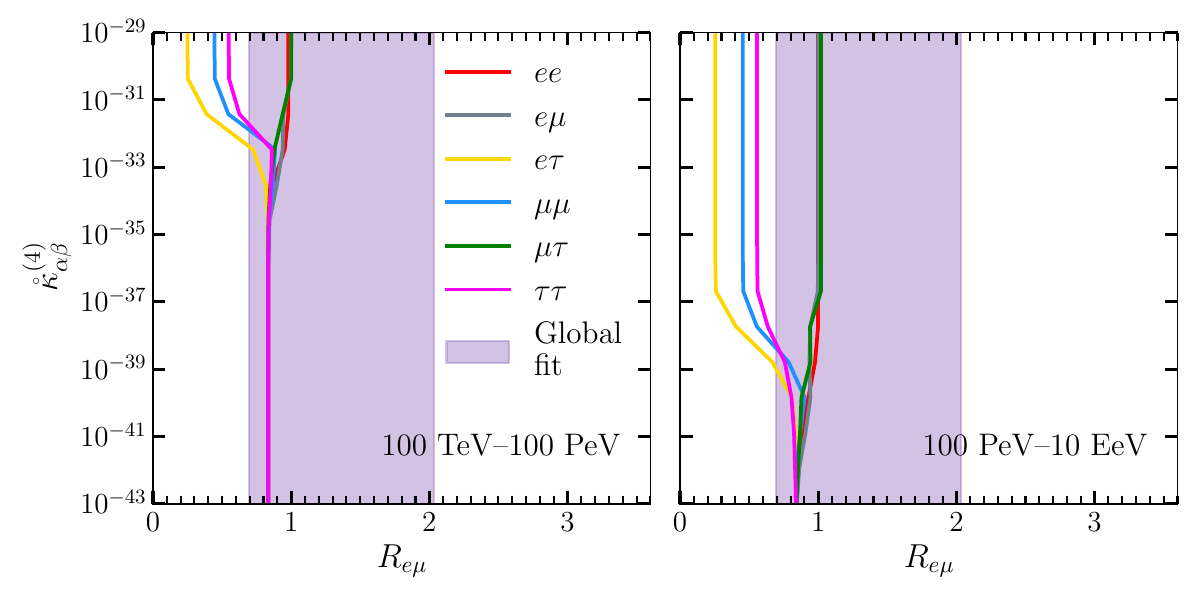}
  \caption{The energy-averaged flavor ratio $R_{e\mu}\equiv \phi_e/\phi_\mu$ for all six flavor combinations of $\mathring{\kappa}^{(4)}_{\alpha\beta}$. In the left (right) panel, we consider $E_\text{min}=100\,\mathrm{TeV}$ and $E_\text{max}=100\,\mathrm{PeV}$ ($E_\text{min}=100\,\mathrm{PeV}$ and $E_\text{max}=10\,\mathrm{EeV}$). We also show the preferred region from the global fit~\cite{Wen:2026ngx} as a purple vertical band.
} 
\label{fig:flux_ratio_2Demu}
\end{figure}

\begin{figure}[h!]
    \centering
       \includegraphics[width=0.5\linewidth]{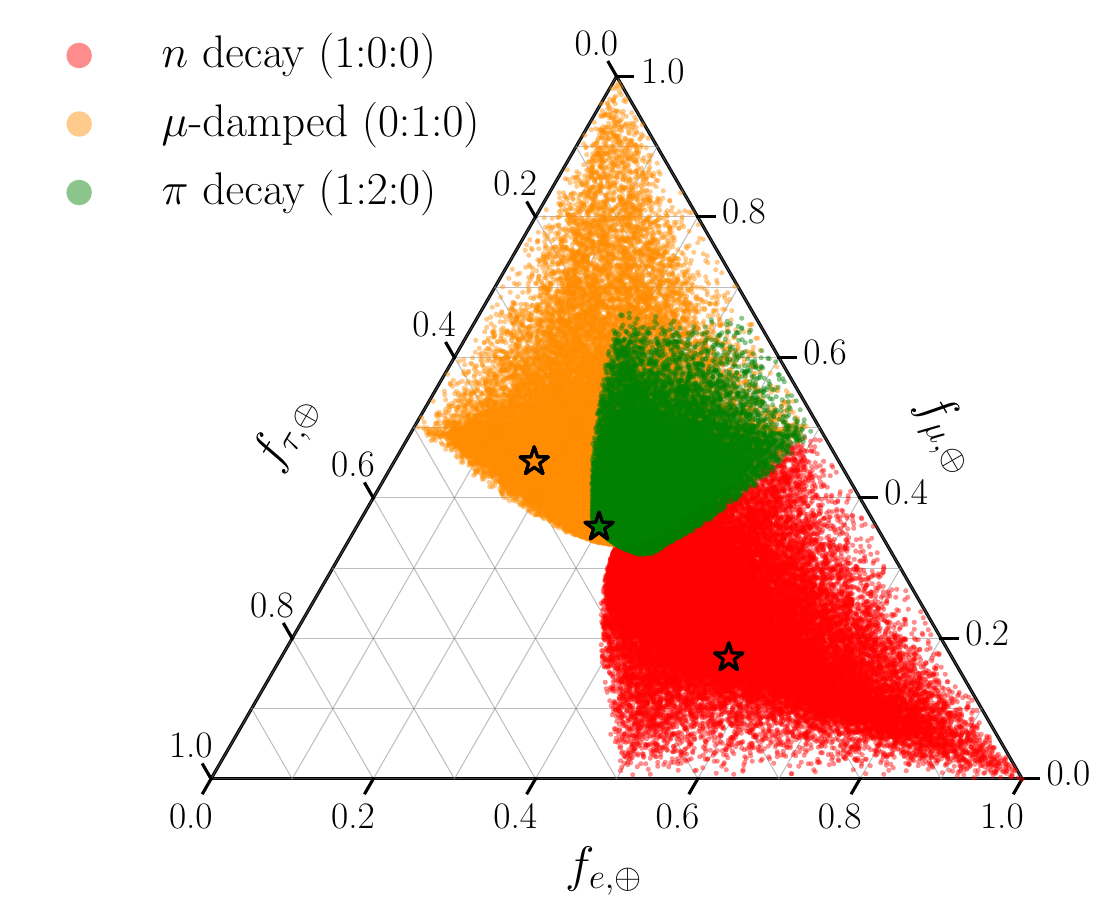}
    \caption{UHE neutrino flavor compositions at Earth in the presence of LIV for pion-decay, muon-damped, and neutron-decay source compositions. The results are obtained from a broad scan over $\mathring{\kappa}^{(4)}_{\alpha\beta}$, allowing all six flavor combinations to be simultaneously nonzero, for neutrino energies between $E_\text{min}=100\,\mathrm{PeV}$ and $E_\text{max}=10\,\mathrm{EeV}$. We use the best-fit NuFIT values for the neutrino mixing parameters. The green region highlights the flavor compositions accessible for cosmogenic neutrinos, with stars indicating the corresponding predictions in the absence of new physics.} 
    \label{fig:ternary_scan}
\end{figure}

\begin{figure*}[t!]
    \centering
    \includegraphics[width=0.49\linewidth]{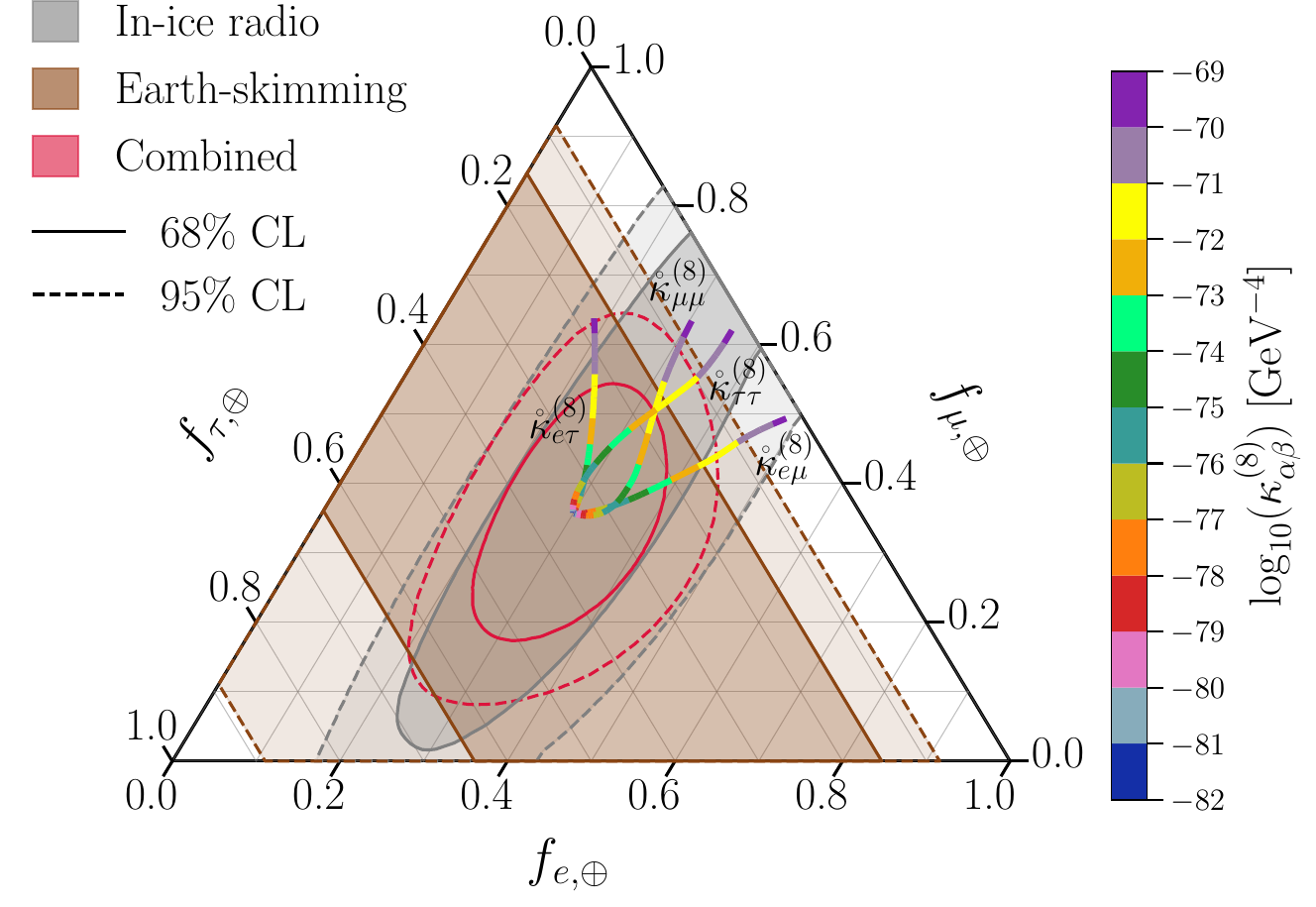}\hfill
    \includegraphics[width=0.49\linewidth]{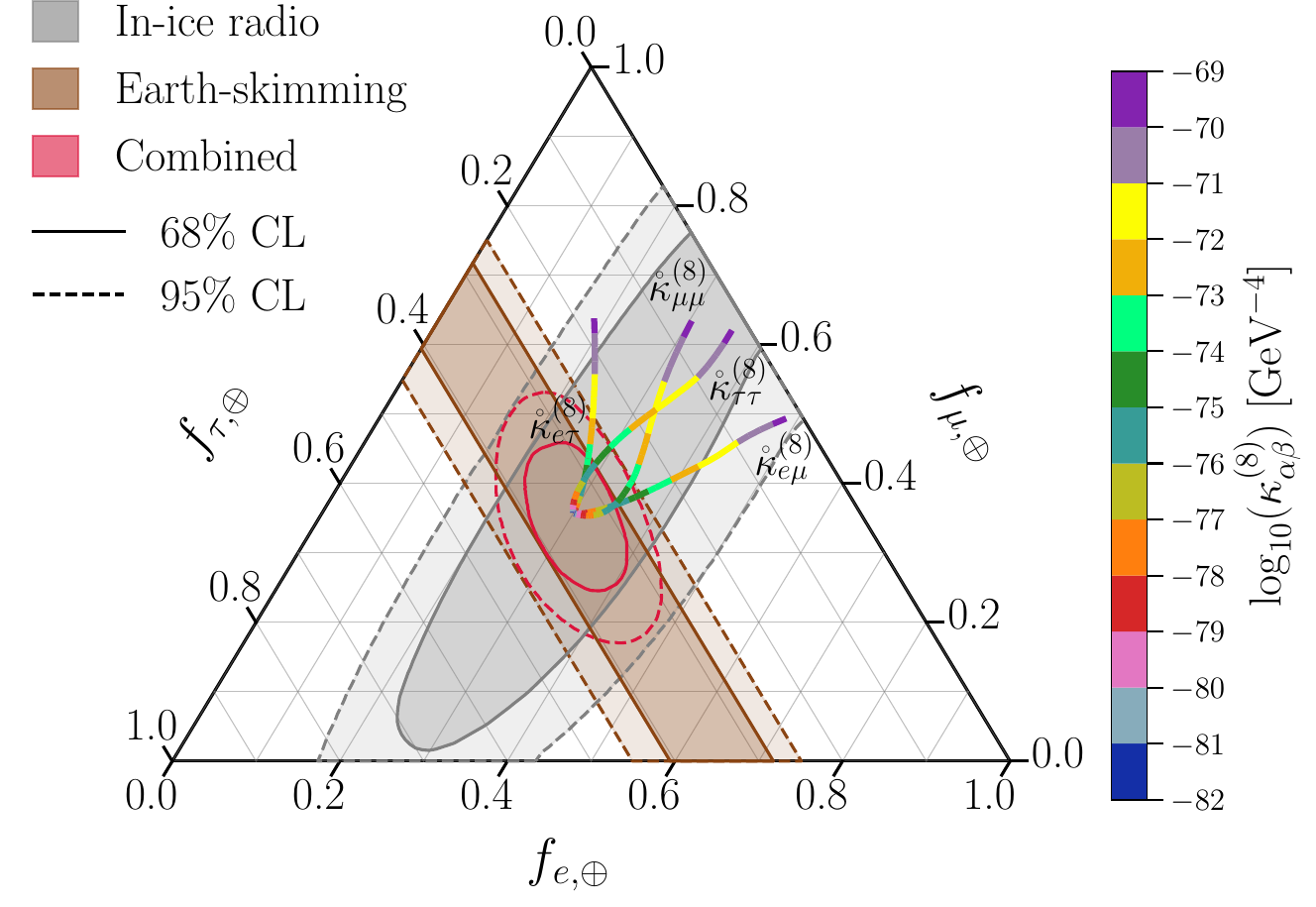}
    \caption{UHE neutrino flavor composition at Earth in the presence of LIV, shown for four different flavor combinations of $\mathring{\kappa}^{(8)}_{\alpha\beta}$, assuming the ``moderate'' flux benchmark.  Also shown are the 68\% and 95\% CL projections for all considered experiments from Ref.~\cite{Burgos-Mondejar:2026ikk} (red), as well as the projections for in-ice radio detection channels (gray) and Earth-skimming tau neutrinos (brown). The left (right) panel shows projections obtained using GRAND10k (GRAND200k) together with all other considered experiments.} 
    \label{fig:ternary_10k_200k}
\end{figure*}

\begin{figure*}[t!]
    \centering
       \includegraphics[width=0.95\linewidth]{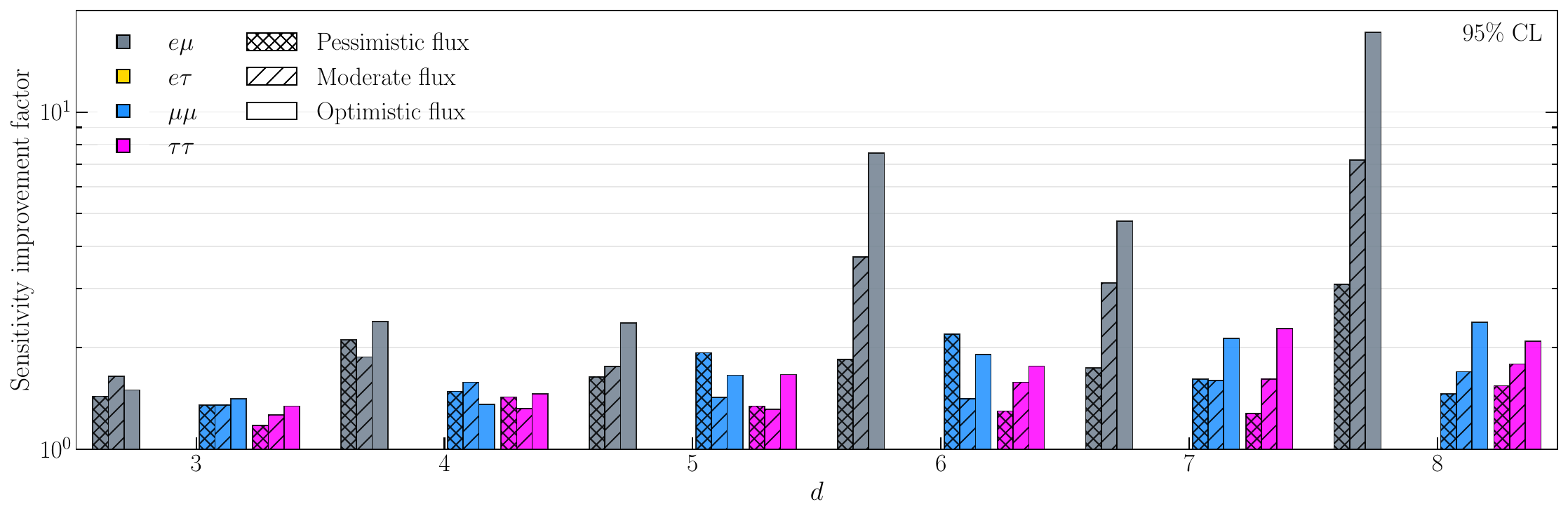}
       \includegraphics[width=0.95\linewidth]{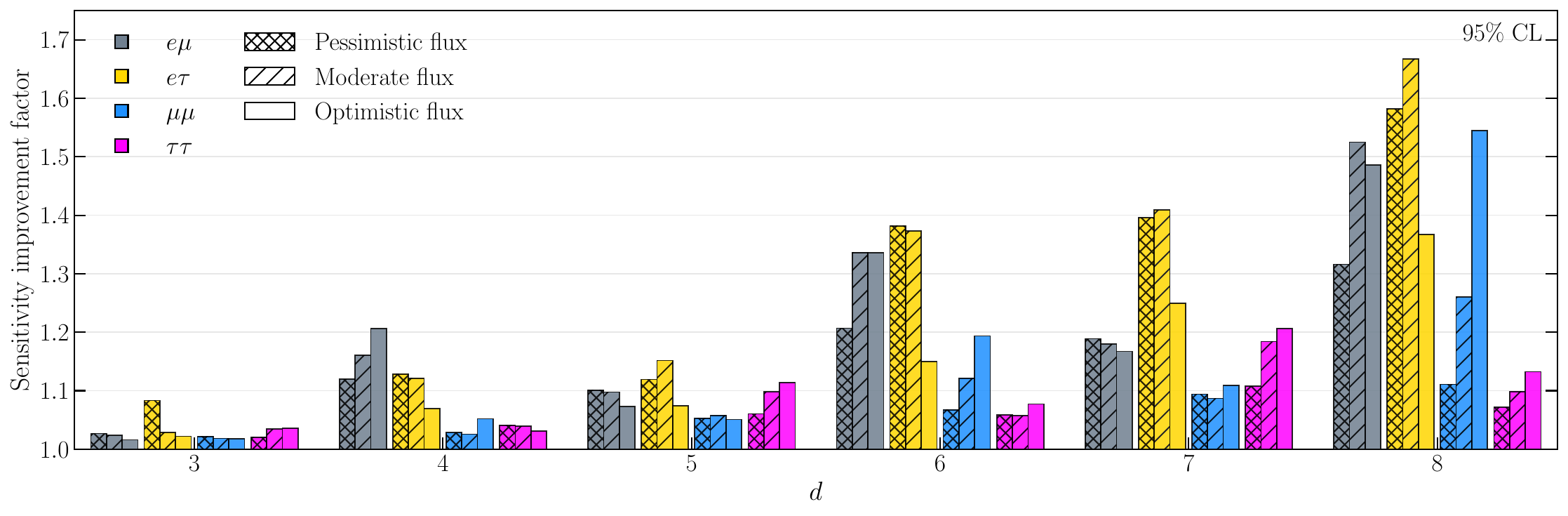}
    \caption{Same as \cref{fig:improvements_full}, but for the GRAND10k (top panel) and GRAND200k (bottom panel) configurations.}
    \label{fig:improvements_10k_200k}
\end{figure*}

\begin{table*}[p]
\centering
\scriptsize
\renewcommand{\arraystretch}{0.95}
\setlength{\tabcolsep}{3.2pt}
\begin{tabular}{ccc cc cc cc}
\hline
& & & \multicolumn{2}{c}{{Pessimistic flux}} & \multicolumn{2}{c}{{Moderate flux}} & \multicolumn{2}{c}{{Optimistic flux}} \\
\cmidrule[0.05em](l{0.5em}r{0.5em}){4-5} \cmidrule[0.05em](l{0.5em}r{0.5em}){6-7} \cmidrule[0.05em](l{0.5em}r{0.5em}){8-9}
$d$ & Parameter & GRAND & Earth-skimming & 3-flavor & Earth-skimming & 3-flavor & Earth-skimming & 3-flavor \\
\hline\hline
\multirow{12}{*}{3} & \multirow{3}{*}{$\mathring{\kappa}^{(3)}_{e\mu}$} & 200k & \sn{1.2}{-30} & \sn{1.1}{-30} & \sn{7.4}{-31} & \sn{7.2}{-31} & \sn{5.3}{-31} & \sn{5.3}{-31} \\
 &  & 50k & \sn{3.3}{-30} & \sn{2.8}{-30} & \sn{1.5}{-30} & \sn{1.3}{-30} & \sn{9.2}{-31} & \sn{8.5}{-31} \\
 &  & 10k & \sn{1.1}{-29} & \sn{7.9}{-30} & \sn{5.7}{-30} & \sn{3.5}{-30} & \sn{2.3}{-30} & \sn{1.5}{-30} \\
\cmidrule[0.03em](l{0.3em}r{0.3em}){2-9}
 & \multirow{3}{*}{$\mathring{\kappa}^{(3)}_{e\tau}$} & 200k & \sn{3.2}{-30} & \sn{3.0}{-30} & \sn{1.3}{-30} & \sn{1.3}{-30} & \sn{7.8}{-31} & \sn{7.6}{-31} \\
 &  & 50k & -- & \sn{1.4}{-29} & \sn{1.2}{-29} & \sn{5.5}{-30} & \sn{2.3}{-30} & \sn{1.8}{-30} \\
 &  & 10k & -- & -- & -- & \sn{1.6}{-29} & -- & \sn{6.1}{-30} \\
\cmidrule[0.03em](l{0.3em}r{0.3em}){2-9}
 & \multirow{3}{*}{$\mathring{\kappa}^{(3)}_{\mu\mu}$} & 200k & \sn{1.1}{-30} & \sn{1.0}{-30} & \sn{6.4}{-31} & \sn{6.3}{-31} & \sn{4.2}{-31} & \sn{4.1}{-31} \\
 &  & 50k & \sn{3.4}{-30} & \sn{3.1}{-30} & \sn{1.5}{-30} & \sn{1.4}{-30} & \sn{8.4}{-31} & \sn{7.9}{-31} \\
 &  & 10k & \sn{2.0}{-29} & \sn{1.4}{-29} & \sn{8.0}{-30} & \sn{5.9}{-30} & \sn{2.7}{-30} & \sn{1.9}{-30} \\
\cmidrule[0.03em](l{0.3em}r{0.3em}){2-9}
 & \multirow{3}{*}{$\mathring{\kappa}^{(3)}_{\tau\tau}$} & 200k & \sn{1.2}{-30} & \sn{1.2}{-30} & \sn{7.1}{-31} & \sn{6.9}{-31} & \sn{4.8}{-31} & \sn{4.6}{-31} \\
 &  & 50k & \sn{3.6}{-30} & \sn{3.3}{-30} & \sn{1.5}{-30} & \sn{1.4}{-30} & \sn{8.9}{-31} & \sn{8.1}{-31} \\
 &  & 10k & \sn{1.2}{-29} & \sn{1.0}{-29} & \sn{6.1}{-30} & \sn{4.8}{-30} & \sn{2.4}{-30} & \sn{1.8}{-30} \\
\hline
\multirow{12}{*}{4} & \multirow{3}{*}{$\mathring{\kappa}^{(4)}_{e\mu}$} & 200k & \sn{2.2}{-39} & \sn{2.0}{-39} & \sn{6.5}{-40} & \sn{5.6}{-40} & \sn{2.2}{-40} & \sn{1.8}{-40} \\
 &  & 50k & \sn{8.5}{-39} & \sn{6.9}{-39} & \sn{3.2}{-39} & \sn{2.3}{-39} & \sn{9.9}{-40} & \sn{6.8}{-40} \\
 &  & 10k & \sn{4.6}{-38} & \sn{2.2}{-38} & \sn{1.5}{-38} & \sn{8.2}{-39} & \sn{6.0}{-39} & \sn{2.5}{-39} \\
\cmidrule[0.03em](l{0.3em}r{0.3em}){2-9}
 & \multirow{3}{*}{$\mathring{\kappa}^{(4)}_{e\tau}$} & 200k & \sn{1.2}{-38} & \sn{1.0}{-38} & \sn{4.5}{-39} & \sn{4.0}{-39} & \sn{1.3}{-39} & \sn{1.2}{-39} \\
 &  & 50k & -- & \sn{7.4}{-38} & \sn{6.8}{-38} & \sn{1.6}{-38} & \sn{9.7}{-39} & \sn{5.8}{-39} \\
 &  & 10k & -- & -- & -- & \sn{8.4}{-38} & -- & \sn{1.5}{-38} \\
\cmidrule[0.03em](l{0.3em}r{0.3em}){2-9}
 & \multirow{3}{*}{$\mathring{\kappa}^{(4)}_{\mu\mu}$} & 200k & \sn{5.5}{-39} & \sn{5.3}{-39} & \sn{2.1}{-39} & \sn{2.0}{-39} & \sn{8.4}{-40} & \sn{8.0}{-40} \\
 &  & 50k & \sn{1.8}{-38} & \sn{1.7}{-38} & \sn{7.8}{-39} & \sn{7.2}{-39} & \sn{3.4}{-39} & \sn{3.0}{-39} \\
 &  & 10k & \sn{1.8}{-37} & \sn{1.2}{-37} & \sn{5.0}{-38} & \sn{3.2}{-38} & \sn{1.4}{-38} & \sn{1.1}{-38} \\
\cmidrule[0.03em](l{0.3em}r{0.3em}){2-9}
 & \multirow{3}{*}{$\mathring{\kappa}^{(4)}_{\tau\tau}$} & 200k & \sn{2.1}{-39} & \sn{2.1}{-39} & \sn{7.2}{-40} & \sn{6.9}{-40} & \sn{3.0}{-40} & \sn{2.9}{-40} \\
 &  & 50k & \sn{8.6}{-39} & \sn{7.8}{-39} & \sn{3.2}{-39} & \sn{2.7}{-39} & \sn{1.1}{-39} & \sn{9.5}{-40} \\
 &  & 10k & \sn{5.1}{-38} & \sn{3.6}{-38} & \sn{1.6}{-38} & \sn{1.2}{-38} & \sn{5.9}{-39} & \sn{4.1}{-39} \\
\hline
\multirow{12}{*}{5} & \multirow{3}{*}{$\mathring{\kappa}^{(5)}_{e\mu}$} & 200k & \sn{1.8}{-47} & \sn{1.7}{-47} & \sn{4.5}{-48} & \sn{4.1}{-48} & \sn{1.4}{-48} & \sn{1.3}{-48} \\
 &  & 50k & \sn{8.3}{-47} & \sn{6.9}{-47} & \sn{2.7}{-47} & \sn{2.1}{-47} & \sn{7.6}{-48} & \sn{5.9}{-48} \\
 &  & 10k & \sn{3.3}{-46} & \sn{2.0}{-46} & \sn{1.4}{-46} & \sn{8.1}{-47} & \sn{5.5}{-47} & \sn{2.3}{-47} \\
\cmidrule[0.03em](l{0.3em}r{0.3em}){2-9}
 & \multirow{3}{*}{$\mathring{\kappa}^{(5)}_{e\tau}$} & 200k & \sn{9.4}{-47} & \sn{8.4}{-47} & \sn{2.9}{-47} & \sn{2.5}{-47} & \sn{6.7}{-48} & \sn{6.2}{-48} \\
 &  & 50k & -- & \sn{4.5}{-46} & \sn{3.3}{-46} & \sn{1.3}{-46} & \sn{7.4}{-47} & \sn{4.3}{-47} \\
 &  & 10k & -- & -- & -- & \sn{6.2}{-46} & -- & \sn{1.3}{-46} \\
\cmidrule[0.03em](l{0.3em}r{0.3em}){2-9}
 & \multirow{3}{*}{$\mathring{\kappa}^{(5)}_{\mu\mu}$} & 200k & \sn{1.7}{-47} & \sn{1.6}{-47} & \sn{3.4}{-48} & \sn{3.2}{-48} & \sn{8.3}{-49} & \sn{7.9}{-49} \\
 &  & 50k & \sn{9.7}{-47} & \sn{8.7}{-47} & \sn{2.6}{-47} & \sn{2.2}{-47} & \sn{5.8}{-48} & \sn{5.0}{-48} \\
 &  & 10k & \sn{1.3}{-45} & \sn{6.9}{-46} & \sn{2.3}{-46} & \sn{1.6}{-46} & \sn{6.4}{-47} & \sn{3.9}{-47} \\
\cmidrule[0.03em](l{0.3em}r{0.3em}){2-9}
 & \multirow{3}{*}{$\mathring{\kappa}^{(5)}_{\tau\tau}$} & 200k & \sn{1.6}{-47} & \sn{1.6}{-47} & \sn{3.3}{-48} & \sn{3.0}{-48} & \sn{8.2}{-49} & \sn{7.4}{-49} \\
 &  & 50k & \sn{8.5}{-47} & \sn{7.7}{-47} & \sn{2.4}{-47} & \sn{2.0}{-47} & \sn{5.6}{-48} & \sn{4.4}{-48} \\
 &  & 10k & \sn{4.2}{-46} & \sn{3.1}{-46} & \sn{1.5}{-46} & \sn{1.1}{-46} & \sn{5.0}{-47} & \sn{3.0}{-47} \\
\hline
\multirow{12}{*}{6} & \multirow{3}{*}{$\mathring{\kappa}^{(6)}_{e\mu}$} & 200k & \sn{2.0}{-56} & \sn{1.6}{-56} & \sn{1.9}{-57} & \sn{1.4}{-57} & \sn{2.9}{-58} & \sn{2.2}{-58} \\
 &  & 50k & \sn{2.7}{-55} & \sn{1.7}{-55} & \sn{3.2}{-56} & \sn{1.6}{-56} & \sn{3.8}{-57} & \sn{1.7}{-57} \\
 &  & 10k & \sn{2.2}{-54} & \sn{1.2}{-54} & \sn{7.1}{-55} & \sn{1.9}{-55} & \sn{9.2}{-56} & \sn{1.2}{-56} \\
\cmidrule[0.03em](l{0.3em}r{0.3em}){2-9}
 & \multirow{3}{*}{$\mathring{\kappa}^{(6)}_{e\tau}$} & 200k & \sn{5.6}{-55} & \sn{4.0}{-55} & \sn{7.3}{-56} & \sn{5.3}{-56} & \sn{7.8}{-57} & \sn{6.7}{-57} \\
 &  & 50k & -- & \sn{2.9}{-54} & \sn{2.9}{-54} & \sn{7.3}{-55} & \sn{3.2}{-55} & \sn{9.3}{-56} \\
 &  & 10k & -- & -- & -- & \sn{3.1}{-54} & -- & \sn{5.5}{-55} \\
\cmidrule[0.03em](l{0.3em}r{0.3em}){2-9}
 & \multirow{3}{*}{$\mathring{\kappa}^{(6)}_{\mu\mu}$} & 200k & \sn{7.6}{-56} & \sn{7.1}{-56} & \sn{9.1}{-57} & \sn{8.1}{-57} & \sn{1.3}{-57} & \sn{1.1}{-57} \\
 &  & 50k & \sn{8.4}{-55} & \sn{7.4}{-55} & \sn{1.5}{-55} & \sn{1.3}{-55} & \sn{2.1}{-56} & \sn{1.5}{-56} \\
 &  & 10k & \sn{1.1}{-53} & \sn{4.9}{-54} & \sn{2.1}{-54} & \sn{1.5}{-54} & \sn{5.3}{-55} & \sn{2.8}{-55} \\
\cmidrule[0.03em](l{0.3em}r{0.3em}){2-9}
 & \multirow{3}{*}{$\mathring{\kappa}^{(6)}_{\tau\tau}$} & 200k & \sn{2.5}{-56} & \sn{2.4}{-56} & \sn{3.4}{-57} & \sn{3.2}{-57} & \sn{6.4}{-58} & \sn{6.0}{-58} \\
 &  & 50k & \sn{2.6}{-55} & \sn{2.2}{-55} & \sn{4.3}{-56} & \sn{3.5}{-56} & \sn{6.8}{-57} & \sn{5.6}{-57} \\
 &  & 10k & \sn{2.2}{-54} & \sn{1.7}{-54} & \sn{6.9}{-55} & \sn{4.4}{-55} & \sn{1.1}{-55} & \sn{6.4}{-56} \\
\hline\hline
\end{tabular}
\caption{Projected 95\% CL sensitivities to the LIV coefficients $\mathring{\kappa}^{(d)}_{\alpha\beta}$ for $d=3$--$8$, considering three flux models and three GRAND configurations (GRAND200k, GRAND50k, and GRAND10k), alongside the other considered experiments. 
We show results for both the Earth-skimming tau-neutrino analysis and the combined three-flavor analysis. All values are given in units of $\mathrm{GeV}^{4-d}$. Dashes indicate parameters that cannot be constrained for the corresponding flux and GRAND configuration.}
\label{tab:liv_sensitivity}
\end{table*}

\begin{table*}[p]
\centering
\addtocounter{table}{-1}
\scriptsize
\renewcommand{\arraystretch}{0.95}
\setlength{\tabcolsep}{3.2pt}
\begin{tabular}{ccc cc cc cc}
\hline
& & & \multicolumn{2}{c}{{Pessimistic flux}} & \multicolumn{2}{c}{{Moderate flux}} & \multicolumn{2}{c}{{Optimistic flux}} \\
\cmidrule[0.05em](l{0.5em}r{0.5em}){4-5} \cmidrule[0.05em](l{0.5em}r{0.5em}){6-7} \cmidrule[0.05em](l{0.5em}r{0.5em}){8-9}
$d$ & Parameter & GRAND & Earth-skimming & 3-flavor & Earth-skimming & 3-flavor & Earth-skimming & 3-flavor \\
\hline\hline
\multirow{12}{*}{7} & \multirow{3}{*}{$\mathring{\kappa}^{(7)}_{e\mu}$} & 200k & \sn{2.0}{-64} & \sn{1.7}{-64} & \sn{1.9}{-65} & \sn{1.6}{-65} & \sn{2.6}{-66} & \sn{2.3}{-66} \\
 &  & 50k & \sn{3.0}{-63} & \sn{2.0}{-63} & \sn{4.0}{-64} & \sn{2.4}{-64} & \sn{4.5}{-65} & \sn{2.6}{-65} \\
 &  & 10k & \sn{2.4}{-62} & \sn{1.4}{-62} & \sn{8.5}{-63} & \sn{2.7}{-63} & \sn{1.3}{-63} & \sn{2.8}{-64} \\
\cmidrule[0.03em](l{0.3em}r{0.3em}){2-9}
 & \multirow{3}{*}{$\mathring{\kappa}^{(7)}_{e\tau}$} & 200k & \sn{4.8}{-63} & \sn{3.5}{-63} & \sn{6.6}{-64} & \sn{4.7}{-64} & \sn{6.9}{-65} & \sn{5.5}{-65} \\
 &  & 50k & -- & \sn{2.9}{-62} & \sn{2.8}{-62} & \sn{7.4}{-63} & \sn{3.1}{-63} & \sn{8.9}{-64} \\
 &  & 10k & -- & -- & -- & \sn{3.2}{-62} & -- & \sn{6.1}{-63} \\
\cmidrule[0.03em](l{0.3em}r{0.3em}){2-9}
 & \multirow{3}{*}{$\mathring{\kappa}^{(7)}_{\mu\mu}$} & 200k & \sn{2.5}{-64} & \sn{2.3}{-64} & \sn{1.9}{-65} & \sn{1.7}{-65} & \sn{2.1}{-66} & \sn{1.9}{-66} \\
 &  & 50k & \sn{4.1}{-63} & \sn{3.4}{-63} & \sn{5.1}{-64} & \sn{3.9}{-64} & \sn{4.6}{-65} & \sn{3.4}{-65} \\
 &  & 10k & \sn{5.9}{-62} & \sn{3.7}{-62} & \sn{1.5}{-62} & \sn{9.3}{-63} & \sn{1.9}{-63} & \sn{8.9}{-64} \\
\cmidrule[0.03em](l{0.3em}r{0.3em}){2-9}
 & \multirow{3}{*}{$\mathring{\kappa}^{(7)}_{\tau\tau}$} & 200k & \sn{1.8}{-64} & \sn{1.6}{-64} & \sn{1.1}{-65} & \sn{9.5}{-66} & \sn{1.1}{-66} & \sn{9.4}{-67} \\
 &  & 50k & \sn{3.0}{-63} & \sn{2.4}{-63} & \sn{3.3}{-64} & \sn{2.4}{-64} & \sn{2.6}{-65} & \sn{1.7}{-65} \\
 &  & 10k & \sn{2.6}{-62} & \sn{2.0}{-62} & \sn{8.2}{-63} & \sn{5.1}{-63} & \sn{1.1}{-63} & \sn{4.8}{-64} \\
\hline
\multirow{12}{*}{8} & \multirow{3}{*}{$\mathring{\kappa}^{(8)}_{e\mu}$} & 200k & \sn{2.1}{-73} & \sn{1.6}{-73} & \sn{7.0}{-75} & \sn{4.6}{-75} & \sn{4.6}{-76} & \sn{3.1}{-76} \\
 &  & 50k & \sn{7.4}{-72} & \sn{3.9}{-72} & \sn{4.0}{-73} & \sn{1.6}{-73} & \sn{1.9}{-74} & \sn{5.8}{-75} \\
 &  & 10k & \sn{1.9}{-70} & \sn{6.3}{-71} & \sn{2.9}{-71} & \sn{4.0}{-72} & \sn{1.7}{-72} & \sn{9.6}{-74} \\
\cmidrule[0.03em](l{0.3em}r{0.3em}){2-9}
 & \multirow{3}{*}{$\mathring{\kappa}^{(8)}_{e\tau}$} & 200k & \sn{2.1}{-71} & \sn{1.3}{-71} & \sn{1.3}{-72} & \sn{8.0}{-73} & \sn{5.3}{-74} & \sn{3.9}{-74} \\
 &  & 50k & -- & \sn{3.2}{-70} & \sn{2.6}{-70} & \sn{3.0}{-71} & \sn{1.1}{-71} & \sn{1.5}{-72} \\
 &  & 10k & -- & -- & -- & \sn{2.7}{-70} & -- & \sn{1.7}{-71} \\
\cmidrule[0.03em](l{0.3em}r{0.3em}){2-9}
 & \multirow{3}{*}{$\mathring{\kappa}^{(8)}_{\mu\mu}$} & 200k & \sn{9.1}{-73} & \sn{8.2}{-73} & \sn{3.1}{-74} & \sn{2.5}{-74} & \sn{1.7}{-75} & \sn{1.1}{-75} \\
 &  & 50k & \sn{3.3}{-71} & \sn{2.6}{-71} & \sn{2.6}{-72} & \sn{1.8}{-72} & \sn{1.0}{-73} & \sn{4.8}{-74} \\
 &  & 10k & \sn{7.8}{-70} & \sn{5.4}{-70} & \sn{1.6}{-70} & \sn{9.2}{-71} & \sn{1.5}{-71} & \sn{6.5}{-72} \\
\cmidrule[0.03em](l{0.3em}r{0.3em}){2-9}
 & \multirow{3}{*}{$\mathring{\kappa}^{(8)}_{\tau\tau}$} & 200k & \sn{3.1}{-73} & \sn{2.9}{-73} & \sn{1.7}{-74} & \sn{1.6}{-74} & \sn{1.6}{-75} & \sn{1.4}{-75} \\
 &  & 50k & \sn{8.5}{-72} & \sn{6.8}{-72} & \sn{6.7}{-73} & \sn{5.2}{-73} & \sn{4.9}{-74} & \sn{3.6}{-74} \\
 &  & 10k & \sn{1.9}{-70} & \sn{1.2}{-70} & \sn{2.9}{-71} & \sn{1.6}{-71} & \sn{2.6}{-72} & \sn{1.3}{-72} \\
\hline\hline
\end{tabular}
\caption{(\textit{Continued.})}
\end{table*}

\newpage

\end{document}